\documentstyle[epsfig]{aipproc}
\begin{document}
\title{Relic Neutrino Asymmetries\footnote{Invited talk at 2nd Tropical
Workshop on Particle Physics and Cosmology, San Juan, Puerto Rico, May 2000.}}
\author{Raymond R Volkas\footnote{r.volkas@physics.unimelb.edu.au}}
\address{School of Physics\\
Research Centre for High Energy Physics\\
The University of Melbourne\\
Victoria 3010 Australia}
\maketitle

\begin{abstract}
I review the topic of relic neutrino asymmetry generation through 
active-sterile
neutrino and antineutrino oscillations in the early universe. Applications to
(i) the suppression of sterile neutrino production, and (ii) the primordial Helium
abundance are briefly presented.
\end{abstract}

Reasonably large relic neutrino asymmetries will be generated by active-sterile
neutrino and antineutrino oscillations in the early universe provided that certain
mild conditions are met \cite{1}. ``Reasonably large'' in this context means
that the
neutrino asymmetry, defined by
\begin{equation}
L_{\nu_{\alpha}} \equiv \frac{n_{\nu_{\alpha}} -
n_{\overline{\nu}_{\alpha}}}{n_{\gamma}},
\end{equation}
where $\alpha = e, \mu, \tau$ and the $n$'s are number densities, can have a final
value as large as about $3/8$ \cite{2,3}. The mild conditions are that the
vacuum mixing angle
$\theta_0$ should be small and that the $\Delta m^2$ should be negative. The
flavour and mass eigenstates are related by
\begin{eqnarray}
| \nu_{\alpha} \rangle & = & \cos\theta_0 | \nu_a \rangle + \sin\theta_0 | \nu_b
\rangle \\ \nonumber
| \nu_{s} \rangle & = & - \sin\theta_0 | \nu_a \rangle + \cos\theta_0 | \nu_b
\rangle,
\end{eqnarray}
so $\Delta m^2 < 0$ means that the predominantly sterile mass eigenstate is lighter
than the predominantly active mass eigenstate. We focus on that epoch of the early
universe subsequent to the disappearance of a significant muon/antimuon component
to the plasma, and ending with the period of Big Bang Nucleosynthesis (BBN).

One reason the generation of neutrino or lepton asymmetries is interesting is
evident from the effective matter potential,
\begin{equation}
V_{eff} = \frac{\Delta m^2}{2p}(- a + b)
\end{equation}
where $p$ is the neutrino momentum, and
\begin{eqnarray}
a & \equiv & - \frac{4 \sqrt{2} \zeta(3)}{\pi^2} \frac{G_F T^3 p}{\Delta m^2}
L^{(\alpha)},\\
b & \equiv & - \frac{4 \sqrt{2} \zeta(3)}{\pi^2} A_{\alpha} \frac{G_F T^4
p^2}{m_W^2 \Delta m^2}.
\end{eqnarray}
The function $a$ is just the generalisation of the usual Wolfenstein effective
potential \cite{4} pertinent to the considered epoch of the early universe.
$G_F$ is the
Fermi constant, $T$ is the temperature and the ``effective asymmetry''
$L^{(\alpha)}$ is given by
\begin{equation}
L^{(\alpha)} \equiv L_{\nu_{\alpha}} + L_{\nu_{e}} + 
L_{\nu_{\mu}} + L_{\nu_{\tau}} + \eta,
\end{equation}
where $\eta \sim 10^{-10}-10^{-9}$ is related to the baryon and electron
asymmetries
required to produce the universe we observe. When lepton asymmetries are large,
the effective potential is also large, leading to small matter-affected
mixing angles and thus also to the suppression of sterile neutrino production.
The asymmetry
generation effect must be taken into account when exploring the implications for
BBN of neutrino scenarios involving light sterile degrees of freedom. In
particular, $\nu_{\mu} \leftrightarrow \nu_s$ oscillation parameters motivated by
the atmospheric neutrino deficit are far from being necessarily ruled out by
BBN \cite{1,2,5,6}. 
The function $b$ is the leading finite-$T$ correction to the Wolfenstein term 
\cite{7}. The
constant $m_W$ is
the $W$-boson mass, while $A_{\alpha}$ is a numerical factor given by $A_e \simeq
17$ and $A_{\mu,\tau} \simeq 4.9$. At high temperatures, $b$ is large enough to
suppress sterile neutrino production by itself. However, it decreases with
temperature as $T^6$, so a large asymmetry is eventually required to keep sterile
neutrino production suppressed at lower temperatures.

Another reason lepton asymmetries are interesting is conveyed by the reactions,
\begin{equation}
\nu_e n \leftrightarrow e^{-} p,\qquad \overline{\nu}_e p \leftrightarrow e^{+} n,
\end{equation}
which play an important role in setting the $n/p$ ratio at weak freeze-out just
prior to BBN. A sufficiently large electron neutrino asymmetry (a few percent) will
alter this ratio and hence change the predicted Helium abundance \cite{3,8,9}.
In particular,
successful BBN can be achieved with a higher than usual baryon density, provided
that $L_{\nu_e}$ is positive and of the correct magnitude. The threatened
overproduction of Helium is compensated by the preferential conversion of neutrons
into protons due to the positive $L_{\nu_e}$.

Light sterile neutrinos are themselves of considerable interest independent of
their possible role in asymmetry generation. For instance, two-fold maximal
mixing is
easily achieved through a pseudo-Dirac structure \cite{10} or through
ordinary-mirror
neutrino mixing \cite{11}. (Mirror neutrinos are strictly speaking not sterile
because they
feel mirror weak interactions. However, they are effectively sterile from the point
of
view of ordinary weak interactions.) Two-fold maximal mixing is of great interest
in light of the solar and atmospheric neutrino deficits. In addition, it is well
known that least one light sterile flavour is required if oscillations are to
simultaneously explain the solar, atmospheric and LSND neutrino data.

As a final comment about motivation, let me also add that
collision and matter affected neutrino oscillation dynamics is a fascinating
subject in its own right. As we will see, neutrino asymmetry generation is driven
by a subtle interplay between quantal coherence and decoherence.

Consider for simplicity a two-flavour active-sterile system $\nu_{\alpha} +
\nu_s$ together with its antiparticle counterpart.
The appropriate dynamical variables are the 1-body reduced density matrices for the
neutrinos and antineutrinos,
\begin{equation}
\rho = \frac{1}{2} ( P_0 + \vec{P} \cdot \sigma ),\quad
\overline{\rho} = \frac{1}{2} ( \overline{P}_0 + \vec{\overline{P}} \cdot \sigma ),
\end{equation}
respectively. The decomposition with respect to the $2 \times 2$ identity matrix
and the Pauli matrices proves to be convenient. All of these quantities are
functions of neutrino momentum $p$ and time $t$ (or equivalently temperature $T$).

The diagonal entries $\rho$ are appropriately normalised distribution functions for
$\nu_{\alpha}$ and $\nu_s$:
\begin{eqnarray}
\rho_{\alpha\alpha} & = & \frac{1}{2}(P_0 + P_z) = \frac{N_{\alpha}}{N^{eq}(0)},
\nonumber \\
\rho_{ss} & = & \frac{1}{2}(P_0 - P_z) = \frac{N_{s}}{N^{eq}(0)},
\end{eqnarray}
where $n_f = \int_0^{\infty} N_f dp$ and $N^{eq}(\xi)$
is the Fermi-Dirac (FD)
distribution with dimensionless chemical potential $\xi \equiv
\mu_{\nu_{\alpha}}/T$,
\begin{equation}
N^{eq}(\xi) = \frac{1}{2\pi^2} \frac{p^2}{\exp(\frac{p}{T} - \xi) + 1}.
\end{equation} 
We have chosen to normalise the neutrino distribution to the 
FD distribution with zero chemical potential.

The off-diagonal entries,
\begin{equation}
\rho_{\alpha s} = \rho_{s\alpha}^{*} = \frac{1}{2} (P_x - i P_y),
\end{equation}
are {\it coherences}. They quantify the amount of quantal or phase coherence
enjoyed
by the neutrinos. These quantities are needed because non-forward neutrino
scattering off the background plasma decreases quantal coherence. The
entries of $\overline{\rho}$ have a similar interpretation.

The evolution of $\rho$ is given by the Quantum Kinetic (or
Boltzmann) Equations (QKEs) \cite{12},
\begin{eqnarray}
\frac{\partial \vec{P}}{\partial t} & = & \vec{V} \times \vec{P} - D \vec{P}_T + 
\frac{\partial P_0}{\partial t} \vec{z},\nonumber \\
\frac{\partial P_0}{\partial t} & \simeq & \Gamma \left[
\frac{N^{eq}(\xi)}{N^{eq}(0)} - \frac{1}{2}(P_0 + P_z) \right],
\end{eqnarray}
where $D = \Gamma/2$ and $\vec{P}_T \equiv P_x \vec{x} + P_y \vec{y}$ with
$\vec{x},\vec{y},\vec{z}$ being unit vectors in the stated directions. The second
equation above has an approximate equality sign because the righthand side assumes
that all background fermions have thermal FD distributions, and that $\nu_{\alpha}$
is distributed in an approximately thermal manner.

The $\vec{V} \times \vec{P}$ term is just a re-expression of coherent
matter-affected oscillatory neutrino evolution. It is equivalent to the usual
Schr\"odinger Equation governing matter-affected neutrino oscillations. Note,
however, that in the early universe this evolution is non-linear because of
neutrino scattering off the background neutrinos of the same flavour. The vector
$\vec{V}$ is given by
\begin{equation}
\vec{V} = \beta \vec{x} + \lambda \vec{z},
\end{equation}
where
\begin{equation}
\beta = \frac{\Delta m^2}{2p} \sin 2\theta_0,\qquad 
\lambda = - \frac{\Delta m^2}{2p} \cos 2\theta_0 + V_{eff}.
\end{equation}
The non-linear effects enter through $V_{eff}$.

The $-D\vec{P}_T$ term drives {\it collisional quantal decoherence}. The
decoherence
rate is equal to half of the collision rate $\Gamma$ for a neutrino of momentum
$p$, where
\begin{equation}
\Gamma \simeq y_{\alpha} G^2_F T^5 \frac{p}{\langle p \rangle}.
\end{equation}
The quantity $\langle p \rangle \simeq 3.15 T$ is the average momentum for a FD
distribution with
zero chemical potential, while $y_e \simeq 4$ and $y_{\mu,\tau} \simeq 2.9$. This
expression for $\Gamma$ is correct provided that thermal equilibrium holds and
asymmetries are not very large. The $-D\vec{P}_T$ term tries to exponentially damp
the coherences $P_{x,y}$ to zero. Since the collision rate goes as $T^5$, quantal
coherence is expunged at high temperatures. In typical applications,
the $\vec{V} \times \vec{P}$ term begins to dominate over the
$-D\vec{P}_T$ term as $T$ approaches a few MeV.

The $\partial P_0/\partial t$ equation describes the repopulation of the
$\nu_{\alpha}$ distribution from the background plasma heat bath. It is
proportional to the total weak collision rate $\Gamma$, and the term in square
brackets on the righthand side acts to drive the actual $\nu_{\alpha}$ distribution
function $(P_0 + P_z)N^{eq}(0)/2$ to FD form $N^{eq}(\xi)$. 

The neutrino asymmetry $L_{\nu_{\alpha}}$ enters these equations through the
function $a$ in $V_{eff}$, and through the chemical potential $\xi$ in $\partial
P_0/\partial t$. For small $L_{\nu_{\alpha}}$,
\begin{equation}
L_{\nu_{\alpha}} \simeq \frac{T^3}{6 n_{\gamma}} \xi.
\end{equation}
In thermal equilibrium the existence of a neutrino asymmetry is equivalent to the
existence of a nonzero $\xi$ (where the chemical potential for antineutrinos is
equal and opposite that of the neutrinos above the chemical
decoupling temperature).

The antineutrino QKEs are of the same form as the above equations with the
substitutions $L_{\nu_{\alpha}} \to - L_{\nu_{\alpha}}$ and $L^{(\alpha)} \to -
L^{(\alpha)}$.

We are primarily interested in the evolution of the asymmetry $L_{\nu_{\alpha}}$.
This is indirectly given by the neutrino and antineutrino QKEs. It is
easy enough, however, to derive a redundant but useful direct evolution equation
for the asymmetry. Employing the QKEs and $\alpha + s$ lepton number conservation
one obtains
\begin{equation}
\frac{d L_{\nu_{\alpha}}}{dt} = \frac{1}{2 n_{\gamma}} \int_0^{\infty} \beta ( P_y
- \overline{P}_y) N^{eq}(0) dp.
\label{dLdt}
\end{equation}
This equation is numerically useful because, during most of its
evolution, $L_{\nu_{\alpha}}$ is the difference of two large numbers (the neutrino
and antineutrino number densities per photon). The use of Eq.(\ref{dLdt})
circumvents this numerical difficulty. This equation is also a very useful starting
point for developing approximate evolution equations \cite{2,13}.

We now discuss the behaviour of the QKEs. We will suppose that the initial
asymmetries are very small or zero, and that a negligible fraction of the
primordial plasma is in the form of sterile states. Also, we consider $|\Delta
m^2|$ values higher than about $10^{-4}$ eV$^2$, because for lower values asymmetry
growth begins when decoherence is negligible, and it tends to be oscillatory
\cite{14}.
 
For sufficiently high temperatures $T$, neutrino and antineutrino oscillations are
severely damped or frozen. There are two reasons for this. First, the
decoherence function $D
\sim T^5$ is large and drives $P_{x,y}$ and $\overline{P}_{x,y}$ to zero, and
hence also
the righthand side of Eq.(\ref{dLdt}) (Quantum Zeno Effect). Second, the $b$-term
in $V_{eff}$ also rises as $T^5$, so the effective matter mixing angle is very
small anyway.

As $T$ decreases with the expansion of the universe, the oscillations begin to
unfreeze. Their initial non-trivial evolution is, however, still dominated by
non-forward scattering. During this phase, a very interesting phenomenon will occur
provided that the mild conditions discussed at the beginning of this article are
met: small $\theta_0$ and $\Delta m^2 < 0$. There is a {\it critical
temperature}
$T_c$, roughly given by
\begin{equation}
T_c \simeq (15 \to 18\ {\rm MeV})\left[ \cos 2\theta_0 \frac{|\Delta m^2|}{{\rm
eV}^2} \right]^{\frac{1}{6}}.
\end{equation}
Above $T_c$, the evolution equations drive $L_{\nu_{\alpha}}$ such as to impel
the effective asymmetry $L^{({\alpha})}$ towards zero (in other words,
$L_{\nu_{\alpha}}$ evolves from zero such as to cancel the $\eta$ term in the
effective asymmetry). We can call this the {\it
asymmetry destruction phase}, and $L^{({\alpha})} = 0$ is a stable fixed point of
the
evolution equations. At $T_c$, this stable fixed point becomes unstable and runaway
positive feedback leads to an exponential increase in $L_{\nu_{\alpha}}$ 
during the {\it explosive growth phase}. The
approximately exponential growth is cut off by the non-linear terms in the QKEs
after $L_{\nu_{\alpha}}$ reaches a value which is several  orders of magnitude
higher than the baryon and electron asymmetries. A {\it power law growth phase}, as
per $L_{\nu_{\alpha}} \sim T^{-4}$, then sets in. Typically, collision
dominated evolution gives way to
oscillation dominated evolution during the power law phase. This is
important, because physically there is a ``hand over'' from collisional
asymmetry growth to non-linear MSW driven growth during this phase. Finally,
asymmetry growth stops when $L_{\nu_{\alpha}}$ reaches about $0.2 - 0.3$.

Some examples of asymmetry evolution are given in Fig.1, which is taken from
Ref.\cite{5}.
These curves were produced by numerically solving the QKEs together with
Eq.(\ref{dLdt}). An analytical understanding of why these curves take the displayed
form has been given in the literature \cite{1,2,3,13}. The key idea is to take
the adiabatic limit
of the QKEs. At high $T$, it is then possible to define precisely what is meant by
collision dominated evolution, and to derive an approximate $d L_{\nu_{\alpha}}/dt$
equation which  makes the critical behaviour at $T_c$ manifest. At low $T$, when
collisions can be neglected, the adiabatic limit of the QKEs is exactly
the same as the adiabatic limit for non-linear MSW evolution. See
Refs.\cite{2,3,13} for a
complete discussion.

\begin{figure}
\epsfig{file=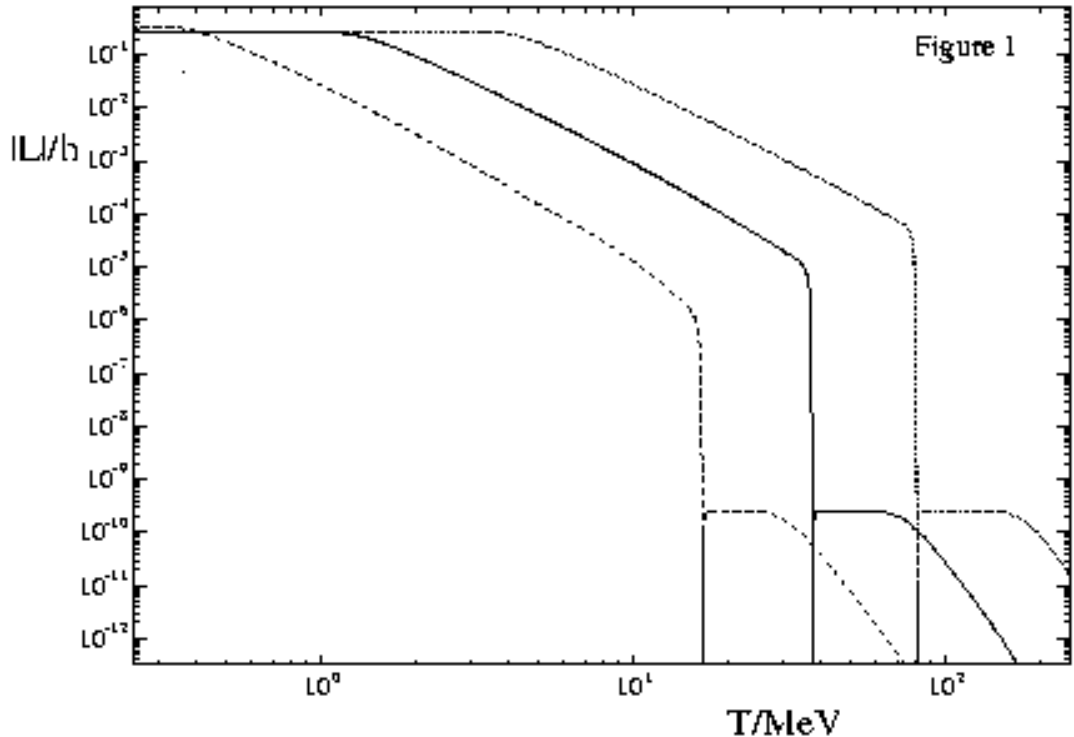,width=15cm}
\caption{Examples of lepton asymmetry growth curves driven by $\nu_{\tau}
\leftrightarrow \nu_s$ and the corresponding antineutrino oscillations. The
mixing angle is selected to be $\sin^2 2\theta = 10^{-8}$. The three curves
correspond to $\Delta m^2 = -0.5,\ -50,\ -5000$ eV$^2$, reading left to
right. 
This figure is taken from Ref.\protect\cite{5}.} 
\end{figure}

I will now review two applications, beginning with the suppression of sterile
neutrino production. Suppose that we wish to solve the atmospheric neutrino problem
by maximal $\nu_{\mu} \to \nu_s$ oscillation with $|\Delta m^2_{\mu s}|$ in the
range $10^{-3}$ eV$^2$ to $10^{-2}$ eV$^2$.\footnote{SuperKamiokande claim that
present data disfavour the $\nu_{\mu} \to \nu_s$ possibility relative to the
$\nu_{\mu}
\to \nu_{\tau}$ mode \cite{15}. However, the dust has yet to settle on this
issue.} If the
$\nu_{\mu} + \nu_s$ subsystem does not mix with any other neutrino, then the
sterile neutrinos (and antineutrinos) will certainly be brought into thermal
equilibrium with the rest of the plasma prior to BBN (unless there are large
pre-existing lepton asymmetries \cite{16}). The resulting increase in the
expansion rate of the universe will lead to a higher than standard Helium-4
abundance. I will assume for the sake of the example that primordial abundance
observations cannot tolerate such an increase in the expansion rate. (A complete
account of the somewhat complicated status of primordial abundance observations
vis-\`a-vis BBN is beyond the scope of this talk.)

Small mixing between the $\nu_s$ and, say, a more massive $\nu_{\tau}$ can
completely change the conclusion that $\nu_s$ is brought into thermal equilibrium.
The $\nu_{\tau}/\nu_s$ oscillation parameters, as chosen in the previous sentence,
satisfy the requirements for large $L_{\nu_{\tau}}$ generation. If the large
$L_{\nu_{\tau}}$ does not have its effects cancelled (see below), and if it is
generated early enough, then the large matter potential consequently generated for
the $\nu_{\mu} \leftrightarrow \nu_s$ mode will suppress these oscillations and
hence also $\nu_s$ production. A numerical calculation is required to properly
analyse the outcome, because the maximally mixed $\nu_{\mu} \leftrightarrow \nu_s$
mode always tends to induce $L^{(\mu)} \to 0$. In other words, the large
$L_{\nu_{\tau}}$ that is being created by the $\nu_\tau \leftrightarrow \nu_s$ mode
could be compensated by $L_{\nu_{\mu}}$ creation driven by $\nu_{\mu}
\leftrightarrow \nu_s$, with the overall effect being that the $\mu$-like effective
asymmetry $L^{(\mu)}$ is driven to zero, and consequently that $\nu_s$ production
through the $\nu_{\mu}$ channel is unsuppressed after all.

Figure 2 shows the outcome of such a numerical calculation \cite{2,5,6}. It is
a plot in
$\nu_{\tau}/\nu_s$ oscillation parameter space. The solid lines correspond to
$|\Delta m^2_{\mu s}| = 10^{-3}, 10^{-2.5}, 10^{-2}$ eV$^2$ in ascending order up
the page. Choose your favourite atmospheric $\Delta m^2_{\mu s}$. In the parameter
region above the relevant solid line, the $L_{\nu_{\tau}}$ asymmetry is {\it not}
cancelled such that $L^{(\mu)} \to 0$. Below the solid line, it is cancelled. The
transition is very sharp. So, above the solid line, $\nu_s$ production via the
$\nu_{\mu} \to \nu_s$ mode is very suppressed. The dot-dashed line refers to
$\nu_s$ production via the other available mode, $\nu_{\tau} \to \nu_s$. This small
angle mode receives very little ``self-suppression'' from its own $L_{\nu_{\tau}}$
creation activity, so its oscillation parameters must satisfy a different sort of
bound. This bound is actually very similar to the old constraints calculated in
Ref.\cite{17} with asymmetry generation artifically switched off. For
illustrative
purposes, the dot-dashed line assumes that no more than $0.6$ extra effective
neutrino flavours are allowed.

\begin{figure}
\epsfig{file=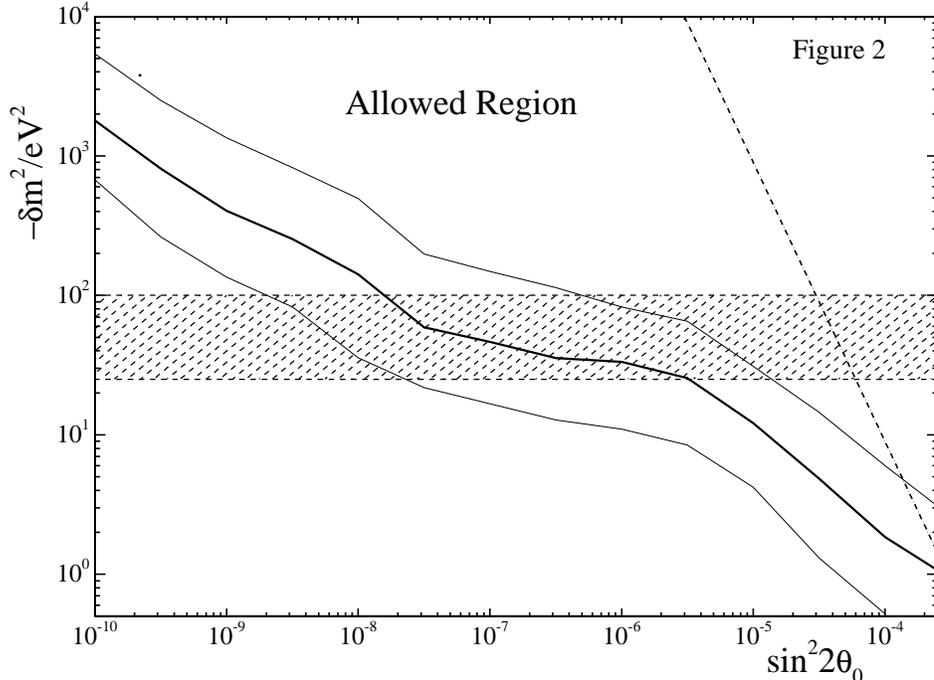,width=12cm}
\caption{Allowed region according to BBN for $\nu_{\tau} \leftrightarrow \nu_s$
oscillation parameters in order to suppress maximal $\nu_{\mu} \leftrightarrow
\nu_s$ oscillations via a large $L_{\nu_{\tau}}$. See text for a complete
discussion. This figure is taken from Ref.\protect\cite{5}.}
\end{figure}

The second application concerns the direct effect of a $\nu_e$ asymmetry on
the neutron to proton ratio at weak freeze out. The implications of neutrino
asymmetry generation in multi-flavour scenarios must be studied on a
case-by-case basis. Here I will consider the scenario of Ref.\cite{18}: there
is one
sterile flavour, which is used to solve the solar neutrino problem via an MSW
style solution. The more massive $\nu_{\mu}$ and $\nu_{\tau}$ states are
maximally mixed, with the mass gap between $(\nu_e,\nu_s)$ and
$(\nu_{\mu},\nu_{\tau})$ set in the LSND range. In this scenario, both the
$\nu_{\mu} \to \nu_s$ and $\nu_{\tau} \to \nu_s$ modes satisfy the conditions
for asymmetry generation. Once $L_{\nu_{\mu}}$ and $L_{\nu_{\tau}}$ have been
generated, these asymmetries can be reprocessed into $L_{\nu_e}$ via
$\nu_{\mu,\tau} \leftrightarrow \nu_e$ oscillations \cite{8}. The $L_{\nu_e}$
outcome
is insensitive to the $\nu_e/\nu_{\mu,\tau}$ mixing angles for a large range
of these parameters because the MSW transitions are adiabatic. It is, however,
somewhat sensitive to $\Delta m^2$ between $(\nu_e,\nu_s)$ and
$(\nu_{\mu},\nu_{\tau})$. Figure 3 shows a plot of the change in the effective
number of neutrino flavours during BBN as a function of this $\Delta m^2$
parameter \cite{8}. The ``effective number of neutrino flavours'' is just a
convenient
measure of the change in the Helium mass fraction $Y_p$ relative to standard
BBN, roughly obeying the relation $\delta Y_p \simeq 0.012 \delta
N_{\nu}^{eff}$. It is affected by both expansion rate alterations and
$L_{\nu_e}$. Figure 3 incorporates both effects.

\begin{figure}
\epsfig{file=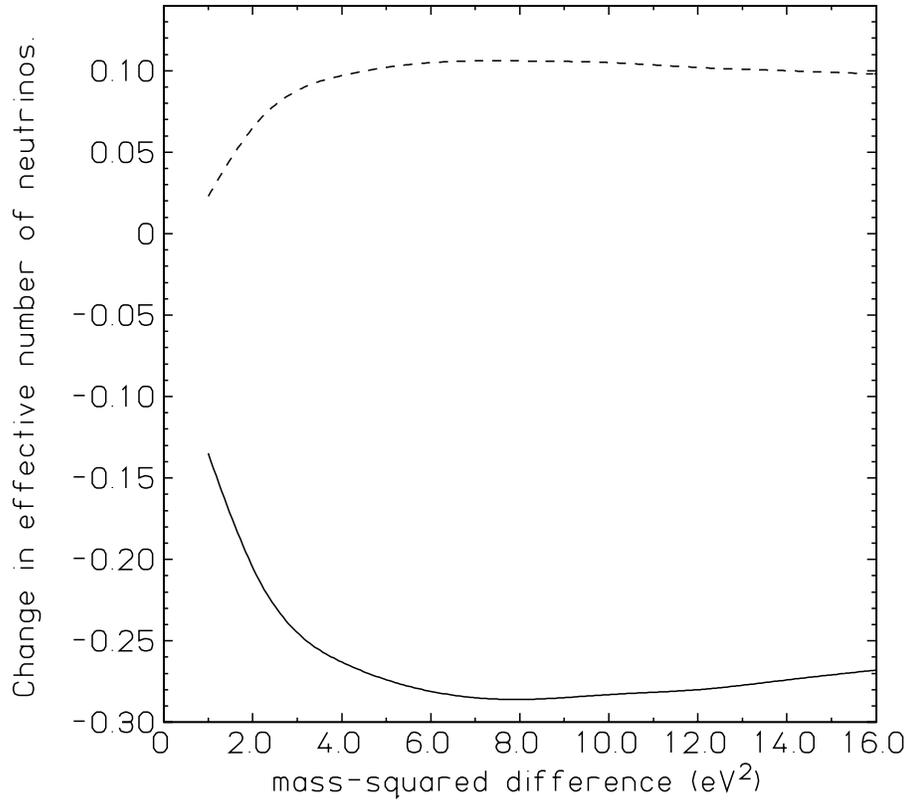,width=12cm}
\vspace{3cm}
\caption{The change in the effective number of neutrino flavours during BBN
for a certain neutrino scenario, plotted as a function of a $\Delta m^2$
parameter that was taken to be in the LSND range. See the text for a more
complete discussion. This figure is taken from Ref.\protect\cite{8}} 
\end{figure}

There are a few other interesting neutrino models that have been analysed in
the literature. A particularly interesting case is the Exact Parity or Mirror
Matter Model \cite{11}. The full analysis is very complicated; please see
Ref.\cite{19} for all
of the details.

To conclude: If light sterile or mirror neutrinos exist, then they should play
a major role in cosmology. The oscillation generated neutrino asymmetry
phenomenon would be a central feature. Sufficiently large neutrino asymmetries
would suppress sterile or mirror neutrino production, and an electron-neutrino
asymmetry of the right magnitude and sign would affect primordial Helium
abundance. In particular, a positive $L_{\nu_e}$ would allow a larger baryon
density to exist without the overproduction of Helium. We note with interest
that the recent Boomerang/Maxima cosmic microwave background anistropy
measurements favour a larger than standard baryon density \cite{20}. While the
combined
solar, atmospheric and LSND data require at least one light sterile neutrino
if oscillations are to simultaneously resolve all of the anomalies, we
especially look forward to future experiments that could provide further
evidence for light sterile neutrinos: SNO, MiniBooNe, the longbaseline
experiments, as well as further SuperKamiokande data.

\acknowledgments{I would like to thank Jose Nieves, Arthur Halprin, Terry
Leung and Qaisar Shafi and all of the very helpful staff for organising an
extremely enjoyable workshop. Thanks to Yvonne Wong for technical assistance.
This work was supported by the Australian
Research Council.}

\end{document}